# Paramagnon Heat Capacity and Anomalous Thermopower in Anisotropic Magnetic Systems: Understanding Inter-Layer Spin Correlations in a Magnetically Disordered Phase

Fatemeh Heydarinasab,[a,*] Morteza Jazandari,[b,*] Md Mobarak Hossain Polash,[c,d,*] Jahanfar Abouie,[b,†] and Daryoosh Vashaee[c,d,†]

[a] Department of Physics, Faculty of Science, University of Sistan and Baluchestan, Zahedan, Iran
[b] Department of Physics, Institute for Advanced Studies in Basic Sciences (IASBS), Zanjan 45137-66731, Iran
[c] Department of Materials Science and Engineering, NC State University, Raleigh, NC 27606, US
[d] Department of Electrical and Computer Engineering, NC State University, Raleigh, NC 27606, US

The interplay between entropy transport and charge carriers-paramagnon interaction in the Onsager linear system has been a subject of debate due to the limited theoretical and experimental understanding of paramagnon heat capacity. In this study, we investigate this interplay in an anisotropic layered magnetic system using cluster mean-field theory with spin quantum correlations. By examining spin correlation functions between different spins with various types of clustering, we derive the spin correlation function as a function of distance and temperature for the inter-layer clusters both below and above the magnetic order phase transition. Our analysis reveals that paramagnons characterized by pronounced spin correlations among inter-layer nearest-neighbor spins exhibit a non-zero heat capacity, providing valuable insights into the dynamics of entropy transport. The findings align with experimental observations, lending strong support to the validity of the paramagnon drag thermopower concept. This study sheds light on the intricate dynamics and thermodynamic properties of paramagnons, advancing our understanding of entropy transport in complex systems.

Spin-driven thermoelectrics has emerged as a promising avenue to enhance the performance of thermoelectric materials. Traditional approaches primarily focused on electron and phonon transport properties, but their limitations have sparked a growing interest in exploring spin-based materials.[1,2,3,4] In particular, the paramagnon-electron drag (PED) thermopower has emerged as a compelling phenomenon, exhibiting continuous enhancement in the paramagnetic phase and offering potential for high-performance thermoelectric devices.[5,6,7]

Paramagnons, arising from thermal fluctuations in the disordered phase, can exhibit magnon-like behavior when specific conditions are met.[5,6] Experimental evidence, including non-zero drag thermopower and extended magnon lifetime, supports the concept.[5,6] The effect has led to achieving a practical zT>1 at T>800 K in MnTe.[6] Furthermore, the magnon-electron drag (MED) phenomenon is linked to the spin-Seebeck effect (SSE) in paramagnetic materials,[8,9,10] while thermal fluctuations enhance zT in $CoSb_3$ super-paramagnets.[11] Ab initio calculations have explored spin correlation functions in the paramagnetic regime, providing a real-space perspective.[12] These findings emphasize the potential of paramagnon-based processes in enhancing thermoelectric performance.

Despite the prospects of PED, the underlying origin of the excess spin-contributed thermopower in paramagnetic materials remains a subject of debate.[5,13] It has been hypothesized that paramagnons, resembling magnons to itinerant carriers, give rise to paramagnon-drag thermopower due to their longer lifetime and larger spatial correlation length compared to free carriers' characteristic length scales.[6] However, a comprehensive understanding of the nature and thermodynamic properties of paramagnons is still lacking. Establishing the paramagnon heat capacity as an essential component for validating the PED thermopower concept has been recognized.[5,13,14]

The heat capacity, which quantifies the entropy change with temperature, and the thermopower, which measures the entropy transported by charge carriers, suggest that paramagnons could contribute to heat capacity through momentum transfer, leading to PED thermopower.[15,16] While experimental observations have shown non-zero magnetic heat capacities above the phase transition temperatures for various magnetic materials,[6,7,17,18] the role of paramagnons in contributing to paramagnetic heat capacity has often been overlooked. Multiple factors, including changes in free energy associated with magnetic ordering-disordering, spin fluctuations, Schottky contributions, and other second-order phase changes, can contribute to non-zero magnetic heat capacities above the transition temperature.[18,19,20,21,22] However, a lack of theoretical understanding of paramagnon heat capacity has hindered a comprehensive assessment of their entropy contribution. In this study, we aim to bridge this gap by conducting a detailed theoretical investigation to understand and quantify the contribution of paramagnons to heat capacity, which can provide insights into the presence of different entropy carriers in the host materials.[18,23,24,25]

To explore the paramagnon heat capacity in disordered spin clusters, we employ a cluster mean-field (CMF) theory, which incorporates thermal fluctuations essential for magnon generation.[5,6,7] Compared to other techniques like quantum Monte Carlo simulations, CMF offers distinct advantages, enabling the examination of short-range orders

---

in different directions and the determination of specific contributions of paramagnon types to the specific heat and thermopower above the critical temperature. In contrast to Monte Carlo simulations, which provide an overall behavior of the specific heat, CMF provides valuable insights into the role of paramagnons in the thermodynamic properties of the system.

We compare the CMF theory with other established approaches commonly used in the study of magnetic materials, such as mean-field (MF) and spin-wave (SW) theories.[26,27] By considering different types of clusters, the CMF theory allows us to investigate the influence of short-range correlations on paramagnon heat capacity and gain a comprehensive understanding of the spin dynamics in paramagnon-electron drag phenomena. To validate our calculations, we compare the derived paramagnon heat capacity with experimental results obtained from MnTe, a material system known for its relevance in studying the magnon/paramagnon drag effect.[5,6,28]

We consider a three-dimensional layered spin-1 system with Heisenberg interactions between nearest-neighbors (NN): $J_1>0$ is the interlayer NN antiferromagnetic (AFM) interaction, $J_2<0$ is the intralayer NN ferromagnetic (FM) interaction, and $J_3>0$ is the diagonal interlayer AFM interaction (Figure 1 (a)). To compare with experimental data, we choose MnTe with $J_1 = 21.5$ K, $J_2 = -0.67$ K, and $J_3 = 2.81$ K.[5,29] The larger value of $J_1$ enables us to categorize the system into two types of layers with different magnetizations ($m_A$ and $m_B$), which we consider in both the MF and CMF calculations. Utilizing these theories, we calculate the staggered magnetization and specific heat by $C_V = \partial \langle H \rangle / \partial T$, where $\langle H \rangle$ is the internal energy and $T$ denotes temperature.[26]

In the MF approximation, the spin-spin interactions are typically replaced by the average of the individual spin components ($S_i$ or $S_j$). This is represented mathematically as $S_i \cdot S_j = S_i \cdot \langle S_j \rangle + S_j \cdot \langle S_i \rangle - \langle S_i \rangle \cdot \langle S_j \rangle$. In our study, we adopt a general approach and consider different sublattices in each layer (as shown in Figure 1 (b)). This allows us to explore a wide range of possibilities and investigate the effects of varying sublattices on the system's behavior. Therefore, the spin Hamiltonian in the MF approximation is written as:

$$H_{MF}/N = 2J_1 \sum_{j=0}^{6} (\vec{S}_{o,j} \cdot \vec{m}_{1,j} + \vec{S}_{1,j} \cdot \vec{m}_{0,j}) + J_2 \sum_{i=0}^{1} \sum_{j,k=0, k \neq j}^{6} \vec{S}_{i,j} \cdot \vec{m}_{i,k} + 2J_3 \sum_{j,k=0, k \neq j}^{6} (\vec{S}_{o,j} \cdot \vec{m}_{1,k} + \vec{S}_{1,j} \cdot \vec{m}_{0,k}) \quad (1)$$

where $\vec{m}_{i,j} = \langle \vec{S}_{i,j} \rangle_{MF}$ is the MF magnetization vector in layer $i$ at site $j$. $N$ is the number of unit cells, wherein each unit cell consists of two layers with seven spins denoted by labels ranging from 0 to 6, as shown in Fig. 1(b), and the summations on $j$ and $k$ run over different sublattices. The self-consistent solution of Eq. (1) yields the MF magnetizations and energy of the system, which, in turn, enables the calculation of the specific heat.

The MF method neglects spin-spin correlations, leading to a lack of finite specific heat in disordered regions. Thus, incorporating spin-spin correlations is crucial for capturing finite specific heat in disordered regions. The Bethe method, considering nearest-neighbor interactions, reveals short-range orders and non-zero specific heat even above the transition temperature, resembling real physical systems.[30] Further modifications, like the Onsager solution, improve consistency with experiments.[30,31] However, applying these methods to our three-dimensional spin system is challenging. To account for larger correlations, we employ CMF theory, treating clusters of multiple sites instead of the single-site approximation.[32,33,34] CMF theory effectively incorporates quantum fluctuations and spin correlations by considering interactions within the clusters and treating spins outside the clusters as effective fields.[35,36,37] Hence, the corresponding CMF Hamiltonian can be defined as:[32,33,34]

$$H_{CMF} = \sum_c (H_c + h_{eff}) \quad (2)$$

where $H_c$ and $h_{eff}$ contain the interactions within and between the clusters, respectively. Here, the summation on c runs over different clusters. For the case of a cluster with four sites, the Hamiltonian, labeled as CMF4, is given by $H_{CMF4} = \sum_c (H_{c-4} + h_{eff-4})$, with

$H_{c-4} = J_1 (\vec{S}_0 \cdot \vec{S}_2 + \vec{S}_1 \cdot \vec{S}_3) + J_2 (\vec{S}_0 \cdot \vec{S}_1 + \vec{S}_2 \cdot \vec{S}_3) + J_3 (\vec{S}_0 \cdot \vec{S}_3 + \vec{S}_1 \cdot \vec{S}_2)$,

$h_{eff-4} = J_1 (\vec{S}_0 \cdot \vec{m}_2 + \vec{S}_1 \cdot \vec{m}_3 + \vec{S}_2 \cdot \vec{m}_0 + \vec{S}_3 \cdot \vec{m}_1)$
$\quad + J_2 \{\vec{S}_0 \cdot (2\vec{m}_0 + 3\vec{m}_1)$
$\quad + \vec{S}_1 \cdot (2\vec{m}_1 + 3\vec{m}_0) + \vec{S}_2 \cdot (2\vec{m}_2 + 3\vec{m}_3)$
$\quad + \vec{S}_3 \cdot (2\vec{m}_3 + 3\vec{m}_2)\}$
$\quad + J_3 \{\vec{S}_0 \cdot (4\vec{m}_2 + 7\vec{m}_3)$
$\quad + \vec{S}_1 \cdot (4\vec{m}_3 + 7\vec{m}_2) + \vec{S}_2 \cdot (4\vec{m}_0 + 7\vec{m}_1)$
$\quad + \vec{S}_3 \cdot (4\vec{m}_1 + 7\vec{m}_0)\}$,

where four-site clusters are illustrated in Fig. 1(c). Here, $\vec{m}_i$ is the magnetization vector at $i^{th}$ sublattice. Like the MF, the self-consistent solution of the CMF Hamiltonian provides the CMF energy and hence the specific heat.

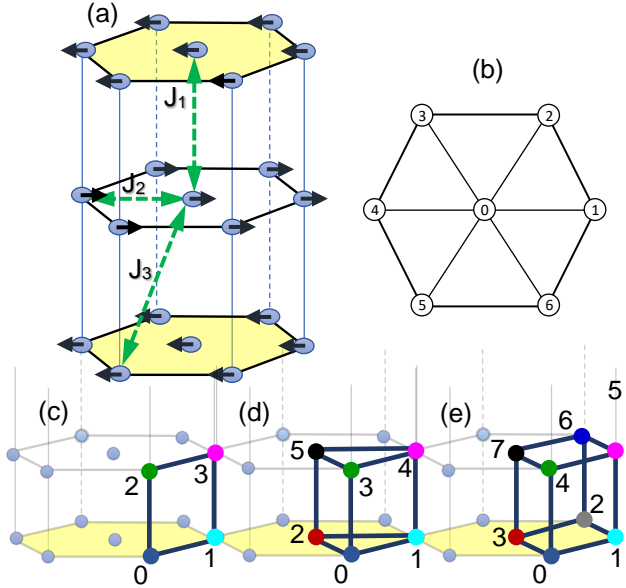

Figure 1: (a) A three dimensional layered spin system with AFM interlayer interactions $J_1$ and $J_3$, and intralayer FM interaction $J_2$. Arrows show the spin order in the AFM phase below a Néel temperature. (b) MF sublattice structure of the spin Hamiltonian in every layer, which respects the symmetry of the lattice. (c), (d) and (e) are four, six, and eight-site clusters used in CMF4, CMF6, and CMF8 theory.

Incorporating the FM $J_2$ interaction, we construct cluster configurations that consider the influence of $J_1$ and $J_3$ interactions (Figure 1(c)-(e)). The corresponding CMF Hamiltonian is defined for each configuration. For an n-site cluster with $n = 4, 6, 8$, the Hamiltonian, denoted as CMFn, is articulated in Eq. (2), with $H_c = H_{c-n}$ and $h_{eff} = h_{eff-n}$ (See the Appendix). Thermal fluctuations are accounted for in MF, SW, and CMF theories, resulting in a reduction of the order parameter until the transition temperature $T_N$, where it eventually vanishes. The CMF theory predicts a lower $T_N$ compared to the MF approximation due to the inclusion of spin correlations (See the Appendix).

The estimated specific heat using the MF approximation, SW theory, and CMF method is illustrated in Figure 2. According to the MF approximation, the $C_V$ grows up to $T_N$, drops sharply at $T_N$, and vanishes at higher temperatures. Compared to the other results, the MF method overestimated the transition temperature. Conversely, the SW method, despite partially including quantum fluctuations, displays a deviation in the specific heat trend over a significant temperature range. Although it provides accurate estimates at low temperatures, it overestimates and saturates as temperature increases. Finally, the specific heat from our CMF(4,6,8) theory, shows a similar trend to the MF method up to the transition temperature. However, unlike the MF approximation, our CMF theory predicts a finite value for the specific heat in the disordered paramagnetic phase. Notably, the CMF theory provides a better estimation of specific heat in both ordered and disordered phases. Estimating non-zero specific heat in disordered phases from the correlations of the spins inside clusters is a successful demonstration of paramagnon contributions. The results can be an essential argument for the existence of the PED thermopower.

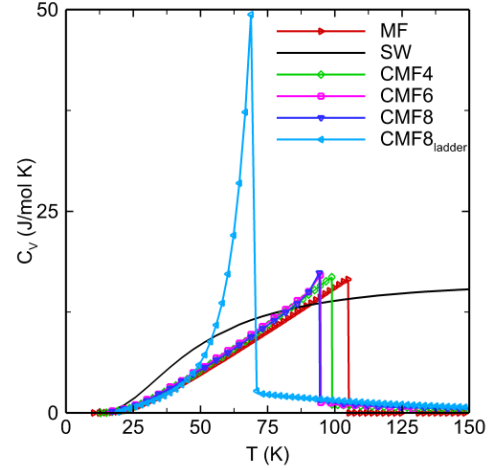

Figure 2: Specific heat estimation for $S = 1$ spin system using MF, SW, CMF4, CMF6, and CMF8 methods. Cluster shapes for CMF4, CMF6, and CMF8 methods are shown in Figure 1. CMF8$_{ladder}$ is the CMF8 results for specific heat with ladder-type configurations presented in the inset of Figure 3 (b). Calculated data is represented by symbols, and solid lines serve as visual guides.

As depicted in Figure 2, considering larger cluster sizes to include more correlations improves the behavior of $C_V$ in the disordered phase, attributed to higher energy fluctuations. Due to the anisotropic nature of our spin system, similar to the ordered phase where two layers host two types of magnons with the same energy dispersion, the disordered phase can exhibit two types of paramagnons. One type corresponds to collective excitations of spins within cuboid-shaped clusters, as illustrated in Figure 1 (e). These paramagnons disperse across the triangular layers and contribute to the system's internal energy, resulting in a non-zero specific heat above the Néel temperature. The other type corresponds to collective excitations of spins within ladder-shaped clusters, aligned with the system's dominant exchange interaction ($J_1$), as shown in the inset of Figure 3 (a). Considering the influence of these paramagnons can lead to intriguing outcomes and provide further insight into the system's behavior.

The strong spin fluctuations in the system are best examined by considering vertical clusters, and including more correlations in the system enhances the observation of the specific heat behavior above the critical temperature (see Figure 2, CMF8$_{ladder}$). Consequently, incrementally increasing the cluster sizes and configurations enhances the paramagnetic specific heat trend towards the transition temperature.

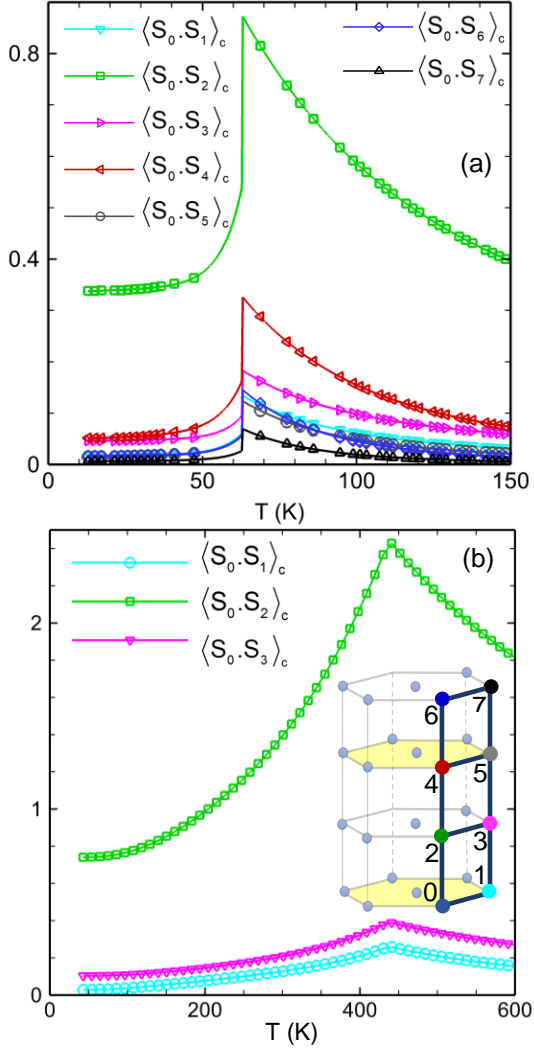

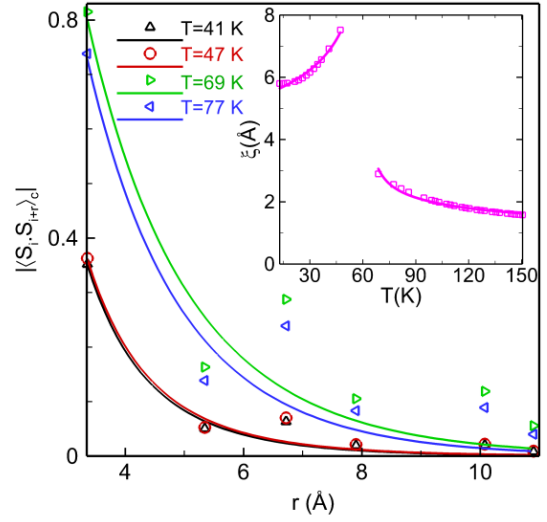

emergence of robust spin excitations, potentially manifesting as paramagnons with the potential for high PED.

In order to find the correlation length and critical exponents, we fitted the absolute value of spin correlation functions with the well-known function of Eq. (4)[30] and plotted its behavior against distance $r$ in Figure 4 for different temperatures (41 K and 47 K for $T < T_N$, and 69 K and 77 K for $T > T_N$).

Figure 4: The absolute value of the spin correlation function as a function of distance ($r$) for various temperatures in the spin-1 system on the MnTe lattice, obtained using the CMF8 approach with ladder-type clustering. The inset depicts the correlation length ($\xi$) versus temperature for the spin-1 system with eight-site clusters and ladder-type configuration, where the symbols represent the calculated results and the solid lines represent the fitted results.

$$\langle S_i . S_{i+r} \rangle_c = A_- \frac{e^{-2r/\xi}}{(r/\xi)^2}, \qquad T < T_N$$
$$= A_+ \frac{e^{-r/\xi}}{(r/\xi)^{1/2}}, \qquad T > T_N \qquad (4)$$

The coefficients $A_\pm$ vary for different temperatures. The critical exponents are calculated from the fit of correlation length with the temperature:[38]

$$\xi = B_- \frac{1}{(1 - T/T_N)^{\nu_-}}, \qquad T < T_N$$
$$= B_+ \frac{1}{(T/T_N - 1)^{\nu_+}}, \qquad T > T_N \qquad (5)$$

which are $\nu_\pm = 0.23$ and $B_+/B_- = 0.32$ for CMF8$_{ladder}$ spin-1 system.

In the context of AFM materials like MnTe, our results are applied to examine the spin correlation functions. MnTe, a spin-5/2 system with a magnetic lattice structure similar to Figure 1(a), shows the estimated spin correlation functions between different sites in the CMF theory, as illustrated in Figure 3(b). The substantial AFM interactions

Figure 3: (a) Spin correlation functions between different spins in the spin-1 system obtained using CMF8 with various types of clustering, as shown in the inset (b). (b) Spin correlation functions between different spins in the $S = 5/2$ system such as MnTe. The symbols represent the calculated results, while solid lines serve as visual guides.

We also estimate the spin correlation functions between NN and next nearest neighbor (NNN) spins along the $J_1$ direction (Figure 3(a)), obtained by CMF8 with a ladder (Figure 3(b)-inset) type of clustering (See the Appendix for the exact definition of this new type of clustering). Here, the spin-spin correlation is defined by:

$$\langle S_i . S_j \rangle_c = \langle S_i . S_j \rangle - \langle S_i \rangle . \langle S_j \rangle \qquad (3)$$

Intriguingly, spin-spin correlations along the J1 direction involving further neighbors (e.g., $\langle S_0 . S_4 \rangle_c$) prove to be stronger than those with direct $J_2$ and $J_3$ interactions (e.g., $\langle S_0 . S_1 \rangle_c$ and $\langle S_0 . S_3 \rangle_c$). These pronounced inter-layer correlations along $J_1$ improve the specific heat trend in the disordered phase, as evident in the CMF8$_{ladder}$ plot in Figure 2. Given this high degree of correlation, selecting clusters along the vertical or $J_1$ direction leads to a pronounced peak, outperforming other clusters. This observation suggests the in MnTe likely account for the observed specific heat behavior (Figure 5), and they are anticipated to have a significant impact on the PED thermopower. Notably, a recent study with neutron-scattering measurements also highlights the presence of remnant AFM correlations at high temperatures.[39] Furthermore, it is evident that the larger correlations between the interlayer nearest-neighbors in the spin-5/2 system can extend into the disordered phase (Figure 3(b), green curve).

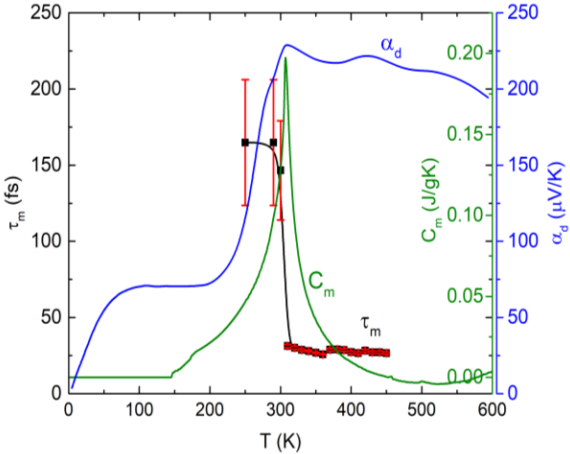

Figure 5: Experimental evidence of the existence of paramagnons in MnTe: magnon/paramagnon lifetime ($\tau_m$) with error-bar from inelastic neutron scattering, magnetic specific heat ($C_m$) extracted from experimental specific heat, and drag thermopower ($\alpha_d$) calculated by subtracting the electronic contribution from total thermopower ($\alpha$).

In magnetic materials, the magnetic specific heat can arise from spin-based entropy carriers like magnons, spin fluctuations, and Schottky contributions, in addition to non-magnetic specific heat contributions from phonons, electrons, and dilation.[5,6,18,24,40] In MnTe, the dominant contribution to the magnetic specific heat stems mainly from magnons, exhibiting a distinct behavior with a sharp peak at $T_N \approx 307$ K and a gradual decrement up to ~600 K, contrary to the typical sharp drop observed in many magnetic materials (Figure 5).[5,6,18,24] Considering the similarities between the spin-1 and spin-5/2 systems, we expect the specific heat trend in the spin-5/2 system to be analogous to that observed in the spin-1 system (Figure 2), which corroborates with the experimental data of MnTe. This suggests the potential significance of paramagnons in contributing to the thermoelectric properties in the disordered phase and supports the existence of the PED thermopower.

With computational limitations in simulating large clusters for the spin-5/2 system, we acknowledge that the specific heat estimation for MnTe is yet to be fully explored. However, based on the trend observed in the specific heat behavior of the spin-1 system (Figure 2), it is reasonable to anticipate similar trends in the specific heat for the spin-5/2 system, providing valuable insights into the thermodynamic properties and the role of paramagnons in enhancing thermoelectric performance.

**Discussion**

The magnitude of the correlation function between spins decreases as the distance between them increases. Notably, near the critical temperature (e.g. ~70 K for spin-one system of Figure 2), the correlation function is larger compared to temperatures further away from the critical temperature. The fluctuating absolute value of the correlation function is attributed to changes in the exchange interaction type within the cluster as the distance increases. As expected, the correlation length rises, forming a cusp near the critical temperature. In the case of a FM system, this cusp leads to a divergence. However, for an AFM interaction, the peak magnitude remains finite. These observations indicate that the most significant correlations and spin fluctuations occur at the critical temperature.

The temperature dependence of the specific heat is a valuable characteristic that aids in the analysis of experimental data. This behavior can also be derived for a two-spin system at high temperatures by employing a high-temperature expansion method, which reveals a $T^{-2}$ dependency in the specific heat. This power-law behavior is not limited to specific spin values but holds true for various spin systems. Comparing the specific heat data of spin-5/2 and spin-1 systems demonstrates that the ratio $C_{5/2}/C_1$ is approximately 20,[‡] suggesting that this ratio remains consistent at high temperatures. Thus, the specific heat of the spin-1 system can serve as an estimate for that of the spin-5/2 in the paramagnetic domain, providing valuable insights for the analysis of experimental data.

In this context, investigating whether paramagnons exhibit quasi-particle behavior with a power-law specific heat as a function of temperature is intriguing. As quantized spin-wave packets arising from thermal fluctuations, paramagnons are expected to contribute significantly to the specific heat due to their association with spin correlations. The persistence of paramagnons at high temperatures suggests quasi-particle-like behavior, leading to the discussed power-law trend in specific heat. Understanding the nature of paramagnons and their connection to the observed power-law behavior requires further theoretical analyses, offering insights into the system's dynamics and the role of short-range order in determining thermodynamic properties.

By considering correlations among spins within clusters that possess a sufficient spin-spin correlation length, we can attribute the paramagnon specific heat to these collective excitations. Notably, when there is a significant disparity in the magnetic interactions between nearest neighbors, the spin-spin correlation becomes more pronounced along the direction of the dominant magnetic interaction above the transition temperature. These enhanced correlations give rise to a non-zero paramagnon specific heat, which in turn supports the observation of the PED thermopower in the disordered phase.

---

[‡] For a two-spin system, the high temperature specific heat can be approximately fit by the functional form $C = 1.25 S^{3.3}$, where $S$ represents the spin value.

In conclusion, the application of CMF theories has significantly advanced our understanding of correlation functions, correlation lengths, and magnetic properties in various systems. By identifying the contributions of different paramagnon types, which represent collective excitations in distinct clusters, to the specific heat, we have gained valuable insights into thermodynamic behavior. These findings underscore the interplay between cluster size, spin-spin correlation length, and magnetic interactions, highlighting their importance in elucidating the thermodynamic properties of magnetic materials. The use of CMF theories has proven instrumental in unraveling these relationships and providing deeper insights into the underlying physics governing the behavior of anisotropic magnetic systems.

**Acknowledgements**

We acknowledge Dr. Mohammad Alidoosti for his assistance in running the codes on the supercomputer. This study is partially based upon work supported by the National Science Foundation (NSF) under grant number CBET-2110603.

**APPENDIX**

**S1. Hamiltonian Formulation within CMF Theory**

In this section, we present the CMF Hamiltonian, employing the CMF theory with clusters containing 6 and 8 sites. Specifically, for the case of a cluster with six sites, the Hamiltonian, denoted as CMF6, is articulated in Eq. (2), where:

$$H_c = H_{c-6}$$
$$= J_1 (\vec{S}_0.\vec{S}_3 + \vec{S}_1.\vec{S}_4 + \vec{S}_2.\vec{S}_5)$$
$$+ J_2 \{\vec{S}_0.\vec{S}_1 + \vec{S}_0.\vec{S}_2 + \vec{S}_1.\vec{S}_2 + \vec{S}_3.\vec{S}_4 + \vec{S}_3.\vec{S}_5 + \vec{S}_4.\vec{S}_5\}$$
$$+ J_3 \{\vec{S}_0.\vec{S}_4 + \vec{S}_0.\vec{S}_5 + \vec{S}_1.\vec{S}_3 + \vec{S}_1.\vec{S}_5 + \vec{S}_2.\vec{S}_3 + \vec{S}_2.\vec{S}_4\} \quad (a1)$$

$$h_{eff} = h_{eff-6} = J_1 \{\vec{S}_0.\vec{m}_3 + \vec{S}_1.\vec{m}_4 + \vec{S}_2.\vec{m}_5 + \vec{S}_3.\vec{m}_0$$
$$+ \vec{S}_4.\vec{m}_1 + \vec{S}_5.\vec{m}_2\}$$
$$+ 2J_2 \{\vec{S}_0.(\vec{m}_1 + \vec{m}_2) + \vec{S}_1.(\vec{m}_0 + \vec{m}_2)$$
$$+ \vec{S}_2.(\vec{m}_0 + \vec{m}_1)\}$$
$$+ 2J_2 \{\vec{S}_3.(\vec{m}_4 + \vec{m}_5) + \vec{S}_4.(\vec{m}_3 + \vec{m}_5)$$
$$+ \vec{S}_5.(\vec{m}_3 + \vec{m}_4)\}$$
$$+ 5J_3 \{\vec{S}_0.(\vec{m}_4 + \vec{m}_5) + \vec{S}_1.(\vec{m}_3 + \vec{m}_5)$$
$$+ \vec{S}_2.(\vec{m}_3 + \vec{m}_4)\}$$
$$+ 5J_3 \{\vec{S}_3.(\vec{m}_1 + \vec{m}_2) + \vec{S}_4.(\vec{m}_0 + \vec{m}_2)$$
$$+ \vec{S}_5.(\vec{m}_0 + \vec{m}_1)\}, \quad (a2)$$

where six sites clusters are illustrated in Fig. 1(d). Here, $\vec{m}_i$ is the magnetization vector at $i^{th}$ sublattice. For eight sites clusters (see Fig. 1(e)), the different terms of the CMF8 Hamiltonian are given by:

$$H_{c-8} = J_1 (\vec{S}_0.\vec{S}_4 + \vec{S}_1.\vec{S}_5 + \vec{S}_2.\vec{S}_6 + \vec{S}_3.\vec{S}_7)$$
$$+ J_2 \{\vec{S}_0.\vec{S}_1 + \vec{S}_0.\vec{S}_3 + \vec{S}_1.\vec{S}_2 + \vec{S}_1.\vec{S}_3$$
$$+ \vec{S}_2.\vec{S}_3\}$$
$$+ J_2 \{\vec{S}_4.\vec{S}_5 + \vec{S}_4.\vec{S}_7 + \vec{S}_5.\vec{S}_6 + \vec{S}_5.\vec{S}_7$$
$$+ \vec{S}_6.\vec{S}_7\}$$
$$+ J_3 \{\vec{S}_0.\vec{S}_5 + \vec{S}_0.\vec{S}_7 + \vec{S}_1.\vec{S}_4 + \vec{S}_1.\vec{S}_6$$
$$+ \vec{S}_2.\vec{S}_5 + \vec{S}_2.\vec{S}_7 + \vec{S}_3.\vec{S}_4$$
$$+ \vec{S}_3.\vec{S}_6\} \quad (a3)$$

$$h_{eff-8}$$
$$= J_1 \{\vec{S}_0.\vec{m}_4 + \vec{S}_1.\vec{m}_5 + \vec{S}_2.\vec{m}_6 + \vec{S}_3.\vec{m}_7 + \vec{S}_4.\vec{m}_0$$
$$+ \vec{S}_5.\vec{m}_1 + \vec{S}_6.\vec{m}_2 + \vec{S}_7.\vec{m}_3\}$$
$$+ J_2 \{\vec{S}_0.(\vec{m}_1 + 2\vec{m}_2 + \vec{m}_3) + \vec{S}_1.(\vec{m}_0 + \vec{m}_2 + \vec{m}_3)$$
$$+ \vec{S}_2.(2\vec{m}_0 + \vec{m}_1 + \vec{m}_3) + \vec{S}_3.(\vec{m}_0 + \vec{m}_1 + \vec{m}_2)\}$$
$$+ J_2 \{\vec{S}_4.(\vec{m}_5 + 2\vec{m}_6 + \vec{m}_7) + \vec{S}_5.(\vec{m}_4 + \vec{m}_6 + \vec{m}_7)$$
$$+ \vec{S}_6.(2\vec{m}_4 + \vec{m}_5 + \vec{m}_7) + \vec{S}_7.(\vec{m}_4 + \vec{m}_5 + \vec{m}_6)\}$$
$$+ J_3 \{\vec{S}_0.(3\vec{m}_5 + 4\vec{m}_6 + 3\vec{m}_7) + 3\vec{S}_1.(\vec{m}_4 + \vec{m}_6 + \vec{m}_7)$$
$$+ \vec{S}_2.(4\vec{m}_4 + 3\vec{m}_5 + 3\vec{m}_7) + 3\vec{S}_3.(\vec{m}_4 + \vec{m}_5 + \vec{m}_6)\}$$
$$+ J_3 \{\vec{S}_4.(3\vec{m}_1 + 4\vec{m}_2 + 3\vec{m}_3) + 3\vec{S}_5.(\vec{m}_0 + \vec{m}_2 + \vec{m}_3)$$
$$+ \vec{S}_6.(4\vec{m}_0 + 3\vec{m}_1 + 3\vec{m}_3)$$
$$+ 3\vec{S}_7.(\vec{m}_0 + \vec{m}_1$$
$$+ \vec{m}_2)\}. \quad (a4)$$

**S2. The Definition of a New Type of Clustering Approach**

Considering that the AFM nearest neighbor interaction, $J_1$, is a prominent factor in our spin model, employing clusters oriented along this direction is likely to yield more accurate results. By partitioning the MnTe lattice into clusters of 8 sites along the $J_1$ direction (as depicted in Fig. 3(b), inset), the Hamiltonians $H_c$ and $h_{eff}$ can be expressed as follows:

$$H_{c-8\,ladder}$$
$$= J_1 \{\vec{S}_0.\vec{S}_2 + \vec{S}_1.\vec{S}_3 + \vec{S}_2.\vec{S}_4 + \vec{S}_3.\vec{S}_5 + \vec{S}_4.\vec{S}_6 + \vec{S}_5.\vec{S}_7\}$$
$$+ J_2 (\vec{S}_0.\vec{S}_1 + \vec{S}_2.\vec{S}_3 + \vec{S}_4.\vec{S}_5 + \vec{S}_6.\vec{S}_7)$$
$$+ J_3 \{\vec{S}_0.\vec{S}_3 + \vec{S}_1.\vec{S}_2 + \vec{S}_2.\vec{S}_5 + \vec{S}_3.\vec{S}_4 + \vec{S}_4.\vec{S}_7$$
$$+ \vec{S}_5.\vec{S}_6\} \quad (a5)$$

$$\begin{aligned}
h_{eff-8\,ladder} = J_1 &\left(\vec{S}_0 \cdot \vec{m}_6 + \vec{S}_1 \cdot \vec{m}_7 + \vec{S}_6 \cdot \vec{m}_0 + \vec{S}_7 \cdot \vec{m}_1\right) \\
+ J_2 &\{\vec{S}_0 \cdot (2\,\vec{m}_0 + 3\,\vec{m}_1) \\
&+ \vec{S}_1 \cdot (2\,\vec{m}_1 + 3\,\vec{m}_0) \\
&+ \vec{S}_2 \cdot (2\,\vec{m}_2 + 3\,\vec{m}_3) \\
&+ \vec{S}_3 \cdot (2\,\vec{m}_3 + 3\,\vec{m}_2) \\
&+ \vec{S}_4 \cdot (2\,\vec{m}_4 + 3\,\vec{m}_5) \\
&+ \vec{S}_5 \cdot (2\,\vec{m}_5 + 3\,\vec{m}_4) \\
&+ \vec{S}_6 \cdot (2\,\vec{m}_6 + 3\,\vec{m}_7) \\
&+ \vec{S}_7 \cdot (2\,\vec{m}_7 + 3\,\vec{m}_6)\} \\
+ J_3 &\{\vec{S}_0 \cdot (2\,\vec{m}_2 + 3\,\vec{m}_3 + 2\,\vec{m}_6 + 4\,\vec{m}_7) \\
&+ \vec{S}_1 \cdot (2\,\vec{m}_3 + 3\,\vec{m}_2 + 2\,\vec{m}_7 + 4\,\vec{m}_6) \\
&+ \vec{S}_2 \cdot (2\,\vec{m}_0 + 3\,\vec{m}_1 + 2\,\vec{m}_4 + 3\,\vec{m}_5) \\
&+ \vec{S}_3 \cdot (2\,\vec{m}_1 + 3\,\vec{m}_0 + 2\,\vec{m}_5 + 3\,\vec{m}_4) \\
&+ \vec{S}_4 \cdot (2\,\vec{m}_2 + 3\,\vec{m}_3 + 2\,\vec{m}_6 + 3\,\vec{m}_7) \\
&+ \vec{S}_5 \cdot (2\,\vec{m}_3 + 3\,\vec{m}_2 + 2\,\vec{m}_7 + 3\,\vec{m}_6) \\
&+ \vec{S}_6 \cdot (2\,\vec{m}_4 + 3\,\vec{m}_5 + 2\,\vec{m}_0 + 4\,\vec{m}_1) \\
&+ \vec{S}_7 \cdot (2\,\vec{m}_5 + 3\,\vec{m}_4 + 2\,\vec{m}_1 \\
&\quad + 4\,\vec{m}_0)\}. \quad (a6)
\end{aligned}$$

## S3. Staggered Magnetization

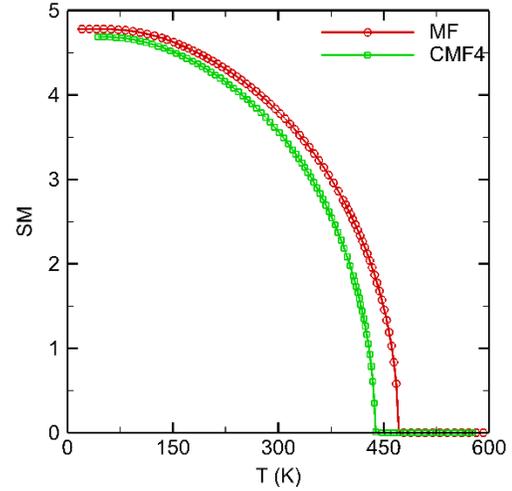

FIG. S1: MF and CMF4 staggered magnetization.